\documentclass[fleqn,usenatbib]{mnras}

\usepackage{newtxtext,newtxmath}
\usepackage{stfloats}

\usepackage[T1]{fontenc}

\DeclareRobustCommand{\VAN}[3]{#2}
\let\VANthebibliography\thebibliography
\def\thebibliography{\DeclareRobustCommand{\VAN}[3]{##3}\VANthebibliography}

\usepackage{graphicx}	
\usepackage{amsmath}	
\usepackage{natbib}

\newcommand{\CII}{C~\textsc{ii}}
\newcommand{\CIV}{C~\textsc{iv}}
\newcommand{\MgII}{Mg~\textsc{ii}}
\newcommand{\SiII}{Si~\textsc{ii}}
\newcommand{\SiIV}{Si~\textsc{iv}}

\title[Ionization evolution of metal absorbers]{E-XQR-30: Evidence for an Increase in the Ionization State of Metal Absorbers from $z\sim$~6 to $z\sim$~2}

\author[S. Rowlands et al.]{
Stephanie Rowlands,$^{1,2}$\thanks{E-mail: stephanierowlands@swin.edu.au}
R. L. Davies,$^{1,2}$\thanks{E-mail: rdavies@swin.edu.au}
E. Ryan-Weber,$^{1,2}$
L. C. Keating$^3$, 
A. M. Sebastian,$^{1,2}$ \newauthor 
G. D. Becker,$^4$
M. Bischetti,$^{5,6}$
S. E. I. Bosman$^{7,8}$,
H. Chen,$^9$ 
F. B. Davies,$^8$ 
V. D'Odorico,$^{6,10}$ \newauthor
P. Gaikwad,$^{11}$ 
S. Gallerani,$^{12}$
M. G. Haehnelt,$^{13,14}$ 
G. Kulkarni,$^{15}$ 
R. A. Meyer,$^{16}$
L. Welsh,$^{17,6}$
Y. Zhu$^{18}$
\\
$^{1}$Centre for Astrophysics and Supercomputing, Swinburne University of Technology, Hawthorn, Victoria 3122, Australia\\
$^{2}$ARC Centre of Excellence for All Sky Astrophysics in 3 Dimensions (ASTRO 3D), Australia\\
$^3$Institute for Astronomy, University of Edinburgh, Blackford Hill, Edinburgh EH9 3HJ, UK \\
$^4$Department of Physics \& Astronomy, University of California, Riverside, CA 92521, USA \\
$^5$Dipartimento di Fisica, Universit\'a di Trieste, Sezione di Astronomia, Via G.B. Tiepolo 11, I-34131 Trieste, Italy \\
$^6$INAF – Osservatorio Astronomico di Trieste, Via G.B. Tiepolo, 11, I-34143 Trieste, Italy \\
$^7$Institute for Theoretical Physics, Heidelberg University, Philosophenweg 12, D–69120, Heidelberg, Germany \\
$^8$Max-Planck-Institut f\"{u}r Astronomie, K\"{o}nigstuhl 17, D-69117 Heidelberg, Germany \\
$^9$Augustana Campus, University of Alberta, Camrose, AB T4V2R3, Canada \\
$^{10}$IFPU – Institute for Fundamental Physics of the Universe, via Beirut 2, I-34151 Trieste, Italy \\
$^{11}$Department of Astronomy, Astrophysics and Space Engineering, Indian Institute of Technology Indore, Simrol, MP 453552, India \\
$^{12}$Scuola Normale Superiore, Piazza dei Cavalieri 7, I-56126 Pisa, Italy \\
$^{13}$Institute of Astronomy, University of Cambridge, Madingley Road, Cambridge CB3 0HA, UK \\
$^{14}$Kavli Institute for Cosmology, University of Cambridge, Madingley Road, Cambridge CB3 0HA, UK \\
$^{15}$Tata Institute of Fundamental Research, Homi Bhabha Road, Mumbai 400005, India \\
$^{16}$Department of Astronomy, University of Geneva, Chemin Pegasi 51, 1290 Versoix, Switzerland \\
$^{17}$Centre for Extragalactic Astronomy, Durham University, South Road, Durham DH1 3LE, UK \\
$^{18}$Steward Observatory, University of Arizona, 933 North Cherry Avenue, Tucson, AZ 85721, USA
}

\date{Accepted XXX. Received YYY; in original form ZZZ}

\pubyear{2024}

\begin{document}
\label{firstpage}
\pagerange{\pageref{firstpage}--\pageref{lastpage}}
\maketitle

\begin{abstract}
We investigate the evolution of the ionization state of metal-enriched gas in and around galaxies near the epoch of reionization using a sample of 488 metal absorption systems at 4.3~$\lesssim z \lesssim$~6.3 from the E-XQR-30 survey. We classify the absorption systems based on whether they display only low-ionization absorption (\CII, \SiII, \MgII), only high-ionization absorption (\CIV, \SiIV), or both. The percentage of low-ionization-only systems decreases from 24\% at $z\sim$~6 to 2\% at $z\sim$~4.3, whilst the fraction of high-ionization-only systems increases from 52\% to 82\%. For mixed absorbers (with both low and high ionization absorption), we use the column density ratios log(N\textsubscript{CII}/N\textsubscript{CIV}) and log(N\textsubscript{SiII}/N\textsubscript{SiIV}) to quantify the average ionization as a function of redshift. The log(N\textsubscript{SiII}/N\textsubscript{SiIV}) ratio does not change significantly over 5~$\lesssim z \lesssim$~6.3. We combine the E-XQR-30 log(N\textsubscript{CII}/N\textsubscript{CIV}) measurements with literature measurements at $z\sim$~2~--~4 and find that the log(N\textsubscript{CII}/N\textsubscript{CIV}) ratio declines by a factor of $\sim$20 between $z\sim$~6 and $z\sim$~2. To explore possible drivers of this evolution, we run photoionization models of gas slabs illuminated by a uniform UV background at fixed density, metallicity and \ion{H}{i} column density. We find that the increase in the ionization state of metal absorbers towards lower redshifts can likely be explained by some combination of 1) an increase in the metallicity of \CIV-absorbing gas and 2) a decrease in the typical H~\textsc{i} column densities of the absorbing gas, driven by the declining cosmic mean density and a rapid rise in the strength of the UV background during the final stages of reionization.
\end{abstract}




\section{Introduction}
The Epoch of Reionization (EoR) is a transitional period impacting almost every baryon, during which the intergalactic medium went from a neutral state opaque to hydrogen ionizing photons to a transparent ionized state \citep{Mesinger_2009, Kollmeier_2014, McQuinn_2016, Caroll_2017}.
Constraining the sources of reionization as well as the exact timescale of when and how rapidly reionization occurred are topics of ongoing research \citep{Becker_Bolton_Lidz_2015, Becker_etal_2015, McQuinn_2016, Fan_2023, Qin_2024}. Neutral hydrogen gas absorbs ionizing photons, preventing direct measurements of the ionizing photon production rate and escape fraction from galaxies prior to the end of reionization. Recently, observations of Ly$\alpha$ emitters with JWST have provided unique constraints on the sizes and distributions of ionized bubbles as far back as $z\simeq$~13, supporting a scenario where reionization is primarily driven by faint, numerous star-forming galaxies \citep[e.g.][]{Mason_2025, Runnholm_2025, Witstok_2025}.

Absorption systems detected against the bright continuum of distant quasars provide a complementary and powerful probe of the properties of the intergalactic and circumgalactic medium during the epoch of reionization \citep{Fan_2006, Becker_Bolton_Lidz_2015, Peroux_2020}. Studies of the Ly$\alpha$ forest and dark gap statistics have revealed that reionization was inhomogeneous and ended late \citep{Becker_2001, Ryden_2017, Zhu_2021, Zhu_2022, Bosman_2022}, with a recent forward modeling study constraining the end of the EoR to $z$~=~5.44~$\pm$~0.02 \citep{Qin_2024}.

Metal absorbers can be detected deep into the EoR and measuring the relative strengths of ions with significantly different ionization potentials offers a direct probe of the UV background (UVB) in the early Universe \citep{Pallottini_2014, Finlator_2016, Doughty_2018}. High ionization species such as C~\textsc{iv} and Si~\textsc{iv} primarily trace warm, diffuse (10\textsuperscript{5}~K) gas in the circumgalactic medium (CGM), whereas neutral and low ionization species such as C~\textsc{ii} and Si~\textsc{ii} typically trace patches of colder (10\textsuperscript{4}~K), denser gas \citep[see][for a review]{Tumlinson_2017}. 

Large surveys of metal absorption systems over cosmic time have revealed that the incidence of high-ionization \CIV~1548,1550\AA\ absorption grows rapidly from $z\sim$~6 to $z\sim$~5 (increasing by a factor of 2~--~4) and then increases more gradually from $z\sim$~5 to $z\sim$~1.5 \citep{Songaila_2001, Becker_2009, RyanWeber_2009, Simcoe_2011, DOdorico_2013, Boksenberg_2015, Bosman_2017, Codoreanu_2018, Meyer_2019, DOdorico_2022, Davies_2023_CIV_paper, Anand_2025}. Similar trends are seen in the \SiIV~1393,1402\AA\ number density \citep{Codoreanu_2018, DOdorico_2022}. In contrast, the incidence of neutral O~\textsc{i}~1302\AA\ and \CII~1334\AA\ absorption decreases from $z\sim$~6 to $z\sim$~5.5 \citep{Becker_2019, Sebastian_2024}, whilst the number densities of medium and weak \MgII~2796,2803\AA\ absorption remain approximately constant \citep{Sebastian_2024}. The metallicity of the CGM increases towards low redshifts as a result of chemical enrichment by star-formation \citep[e.g.][]{Madau_2014, Deepak_2025}. The marked decrease in O~\textsc{i} and \CII\ between $z\sim$~6 and $z\sim$~5.5 thus points towards an increase in the ionization state of the absorbing gas, consistent with the rapid rise in high-ionization absorption over the same redshift range. This increase in ionization state may be driven by an increase in the strength and/or hardness of the ionizing photon background near the end of the EoR \citep{Becker_2019, Davies_2023_CIV_paper}.

The change in ionization state can be directly examined by comparing the column densities of ions with substantially different ionization potentials. \citet{DOdorico_2013} found that the column density ratio of \SiIV\ (with an ionization potential of 33.5~eV) relative to \CIV\ (47.9 eV) decreases from $z\sim$~5 towards low redshift, suggesting an increase in the ionization state of the absorbing gas. However, the N\textsubscript{SiIV}/N\textsubscript{CIV} ratio is also sensitive to variations in the relative abundances of carbon and silicon. Comparing different ions of the same element, such as \CIV\ and \CII\ (11.3~eV), removes this systematic uncertainty. \citet{Cooper_2019} compared the properties of low-ionization and high-ionization absorption in 69 intervening absorption systems at $z~>$~5 observed with Magellan/FIRE and found evidence for a sharp decline in the N\textsubscript{CII}/N\textsubscript{CIV} ratio from $z\sim$~6 to $z\sim$~5.5, which they suggest could be driven by a rapid increase in either the strength/hardness of the UV background or the metallicity of the diffuse \CIV-absorbing gas. However, the majority of the absorption systems studied in \citet{Cooper_2019} were detected in only one of \CII\ or \CIV, meaning that only upper or lower limits on the N\textsubscript{CII}/N\textsubscript{CIV} ratio could be obtained. This makes it difficult to robustly measure how the average ratio changes with redshift.

Here, we further investigate the ionization properties of metal absorption systems near the end of the EoR using a sample of 488 metal absorbers from the enlarged Ultimate XSHOOTER legacy survey of quasars at \mbox{$z\sim$~5.8~--~6.6} (E-XQR-30). This survey obtained deep observations at a spectral resolution of R~$\simeq$~10,000, providing the largest homogeneous catalog of metal absorbers at $z >$~5 \citep{Davies_2023_Catalogue_paper}. The E-XQR-30 catalog is a factor of three deeper than the \citet{Cooper_2019} dataset and contains more than twice as many absorption systems at $z>$~5. The large sample enables us to separately investigate the evolution of systems showing only high-ionization species, only low-ionization species, and systems showing a mix of both. Focusing on \CII, \CIV, \SiII\ and \SiIV, we explore how the fraction of low-ionization, high-ionization and mixed absorption systems varies as a function of redshift. For systems with detections of both low-ionization and high-ionization absorption, we quantify the redshift evolution in the column density ratios N\textsubscript{CII}/N\textsubscript{CIV} and N\textsubscript{SiII}/N\textsubscript{SiIV}. The E-XQR-30 measurements of the carbon and silicon ions span 4.3~$< z <$~6.5 and 4.9~$< z <$~6.5, respectively, covering the tail end of the EoR. We expand the measurements of N\textsubscript{CII}/N\textsubscript{CIV} to lower redshifts by using data for absorption systems at \mbox{2 $< z <$ 4 from \citet{Boksenberg_2015}}. 

We describe the E-XQR-30 survey with the absorber sample and the calculation of column density ratios and absorber fractions in Section \ref{sec:data}. Our results on the redshift evolution of metal absorber properties are presented in Section \ref{sec:results}. We discuss comparisons to \textsc{Cloudy} models and the implications for reionization in Section \ref{sec:modelling} and present our conclusions in Section \ref{sec:conclusions}.

\section{Samples and Data Analysis}
\label{sec:data}
\subsection{E-XQR-30 Sample}
Our primary sample is drawn from the E-XQR-30 metal absorber catalogue \citep{Davies_2023_Catalogue_paper}. E-XQR-30 is a deep spectroscopic survey of 42 of the brightest quasars at $z\sim$~5.8~--~6.6 \citep{DOdorico_2023} with absolute magnitudes of $-27.8~<$~M\textsubscript{1450\AA}~$<-26.2$ \citep{Bischetti_2022}, designed to study quasars, enriched gas and the Lyman-$\alpha$ forest during the first billion years of the Universe. The survey represents the most extensive collection of high-quality spectra of early Universe quasars. The observations were conducted using VLT/X-Shooter which has 3 spectroscopic arms that together cover near-ultraviolet to near-infrared wavelengths from 3000~--~24800\AA\ \citep{Vernet_2011}. The E-XQR-30 spectra have a median spectral resolution of 11400 in the VIS arm and 9800 in the NIR arm and a median SNR of 30 per 10 km/s spectral channel \citep{DOdorico_2023}.

The metal absorption systems were cataloged as described in \citet{Davies_2023_Catalogue_paper}. An initial list of candidate systems was generated by performing an automated search for \MgII, Fe~\textsc{ii}, \CII, \CIV, \SiIV, and/or N~\textsc{v}~1238,1242\AA\ absorption. The candidates were filtered using a combination of automated algorithms and visual inspection. The Ly$\alpha$ forest is saturated at these redshifts, and therefore only the wavelength regions redward of the quasar Ly$\alpha$ emission line were searched. Table~\ref{table:properties_of_ions} shows the complete redshift range over which each ion can theoretically be detected when combining the data from all 42 quasars. Absorption components separated by less than 200~km~s$^{-1}$ were grouped into systems with column density given by the sum of the column densities of the constituent components. The full E-XQR-30 catalog contains 778 absorption systems over a redshift range of 2~$\leq z \leq$~6.5 \citep{Davies_2023_Catalogue_paper}. The E-XQR-30 catalog has a 50\% completeness limit of log(N\textsubscript{CIV}/cm\textsuperscript{-2})~=~13.2, and therefore the absorption systems studied in this work most likely probe CGM gas. \citet{DOdorico_2016} showed that log(N\textsubscript{CIV}/cm\textsuperscript{-2})~$>$~12 corresponds to the CGM of galaxies at $z\sim$~3, and $z\sim$~6 \CIV\ absorbers with log(N\textsubscript{CIV}/cm\textsuperscript{-2})~$>$~13~--~13.5 have associated galaxies within 200~pkpc \citep{Diaz_2021, Kashino_2023}.

\begin{table}
\small
\centering
\begin{tabular}{|p{2.1cm}|p{1.4cm}|p{1.5cm}|p{1.2cm}} 
 \hline
 Ion & Ionization energy (eV) & Redshift Range & Number of systems \\ [0.5ex] 
 \hline
 \CII\ 1334~\AA\ & 11.3 & 5.17~--~6.34 & 21 \\ 
 \CIV\ 1548,1550~\AA\ & 47.9 & 4.34~--~6.34 & 444 \\
 \SiII\ 1526~\AA\ & 8.2 & 4.42~--~6.34 & 21 \\
 \SiIV\ 1393,1402~\AA\ & 33.5 & 4.94~--~6.34 & 85\\
 \MgII\ 2796,2803~\AA\ & 7.6 & 2.04~--~6.34 & 90\\
 \hline 
\end{tabular}
\caption{List of carbon, silicon and magnesium ions used in this study and their respective ionization potential energies \citep{NIST_ASD}, along with the detectable redshift range from the E-XQR-30 spectra and the number of E-XQR-30 absorption systems with detections.}
\label{table:properties_of_ions}
\end{table}

\subsection{Carbon and Silicon Absorber Properties}
To determine whether the ionization state of the CGM changes with redshift, we focus on the singly ionized and triply ionized states of carbon (\CII\ and \CIV) and silicon (\SiII\ and \SiIV). The doubly ionized ions of carbon and silicon, C~\textsc{iii}~977\AA\ and Si~\textsc{iii}~1206\AA, could not be used because they fall in the Ly$\alpha$ forest. Four strong \SiII\ lines fall within the wavelength range of the E-XQR-30 spectra: 1260\AA, 1304\AA, 1526\AA, and 1808\AA. \SiII~1260\AA\ is the strongest line but is only detectable at 5.5~$\lesssim z\lesssim$~6.3. The second strongest line, \SiII~1526\AA, is six times weaker but detectable over a significantly larger redshift range, 4.4~$\lesssim z\lesssim$~6.3. If we were to use both \SiII~1260\AA\ and \SiII~1526\AA\ at $z\gtrsim$~5.5 and only \SiII~1526\AA\ at lower redshifts, the 50\% completeness limit for \SiII\ detection would be 6$\times$ deeper in the highest redshift interval, making it difficult to interpret potential variations in the ionization state of silicon absorbers as a function of redshift. Therefore, we exclusively use \SiII~1526\AA\ because it provides the best compromise between oscillator strength and redshift range.

We select absorption systems containing any of the \CII, \CIV, \SiII\ and \SiIV\ metal absorption lines, with no restrictions in column density. We restrict our sample to systems at redshifts where both ions of either carbon or silicon fall redward of the quasar Ly$\alpha$ line and outside of masked spectral regions \citep{Davies_2023_Catalogue_paper}. This results in a sample of 172 systems (after grouping components within 200~km~s$^{-1}$). We further restrict the sample to intervening absorbers (at least 10,000~km/s below the quasar redshift) to avoid absorption systems potentially impacted by the strong quasar radiation field \citep[e.g.][]{Berg_2016, Perrotta_2016} or heating due to quasar-driven winds \citep[e.g.][]{Bischetti_2022, Bischetti_2023}. This leaves a total sample of 148 systems.

We use the column density ratios log(N\textsubscript{CII}/N\textsubscript{CIV}) and log(N\textsubscript{SiII}/N\textsubscript{SiIV}) as proxies for the ionization state of the absorbing gas. If both ions of silicon or carbon are robustly detected in an absorption system, the column density ratio can be calculated directly. We take the column density measurements of saturated systems at face value, although they have large associated uncertainties. 12 systems have both \CII\ and \CIV\ detected and are classified as “carbon both” in the analysis, and 14 systems have both \SiII\ and \SiIV\ detected and are classified as `silicon both'. For systems in which only one ion is present or either ion is flagged as an upper limit in \citet{Davies_2023_Catalogue_paper}, we calculate a limit on the column density ratio based on the noise in the spectrum, as described in Section \ref{subsec:limit_calculations}. 9 systems are only detected in \CII\ and classified as \CII-only, resulting in a lower limit on the column density ratio. 34 systems have only \CIV\ detected and are similarly classified as \CIV-only, resulting in an uppper limit on the column density ratio. The same applies to the silicon systems. The final absorber sample is summarised in Table~\ref{table:table_of_ions_and_systems}.

\begin{table*}
\small
\centering
\begin{tabular}{c|ccc|ccc|c} 
 \hline
 \multicolumn{1}{c|}{Dataset} & \multicolumn{3}{c|}{$n$ (E-XQR-30)} & \multicolumn{3}{c|}{$n$ (Cooper et al. 2019)} & \multicolumn{1}{c}{$n$ (Boksenberg et al. 2015)}\\ 
 Redshift & 4.3~$<z<$~5.17 & 5.17~$<z<$~5.7 & $z>$~5.7 & 4.3~$<z<$~5.17 & 5.17~$<z<$~5.7 & $z>$~5.7 & 2 $< z <$ 4.4\\
 \hline
 \CII~+~\CIV\ & N/A & 3 & 9 & N/A & 0 & 6 & 22\\ 
 \CII, no \CIV\ & N/A & 4 & 5 & N/A & 2 & 10 & N/A\\
 \CIV, no \CII\ & N/A & 15 & 20 & N/A & 5 & 4 & 39\\
 \hline
 \MgII\ + \CIV\ & 52 & 17 & 6 & 1 & 6 & 0 & N/A\\ 
 \MgII, no \CIV\ & 5 & 11 & 6 & 3 & 4 & 1 & N/A\\
 \CIV, no \MgII\ & 264 & 52 & 13 & 2 & 6 & 0 & N/A\\
 \hline
 \SiII~+~\SiIV\ & 2 & 6 & 6 & 0 & 4 & 4 & 4\\ 
 \SiII, no \SiIV & 0 & 1 & 6 & 1 & 5 & 12 & N/A\\
 \SiIV, no \SiII\ & 15 & 41 & 15 & 1 & 7 & 2 & 10\\
 \hline
\end{tabular}

\caption{The number of absorption systems, denoted as $n$, in the E-XQR-30, \citet{Cooper_2019} and \citet{Boksenberg_2015} samples that have one or both of \CII\ and \CIV, \MgII\ and \CIV, and \SiII\ and \SiIV. `\CIV, no \CII' refers to systems with detections of \CIV\ and not \CII\ within the \CII\ redshift range. These systems overlap with `\CIV, no \MgII' which refers to detections of \CIV\ with no \MgII\ within the \MgII\ redshift range. \MgII\ is not accessible for some \CII\ systems at $z >$~5.7 due to skyline contamination. `N/A' indicates redshift ranges where one or both of the relevant ions are detectable or combinations of ions not probed by the dataset.}
\label{table:table_of_ions_and_systems}
\end{table*}

\begin{figure}
	\includegraphics[clip = True, trim = 10 0 40 40, width=\columnwidth]{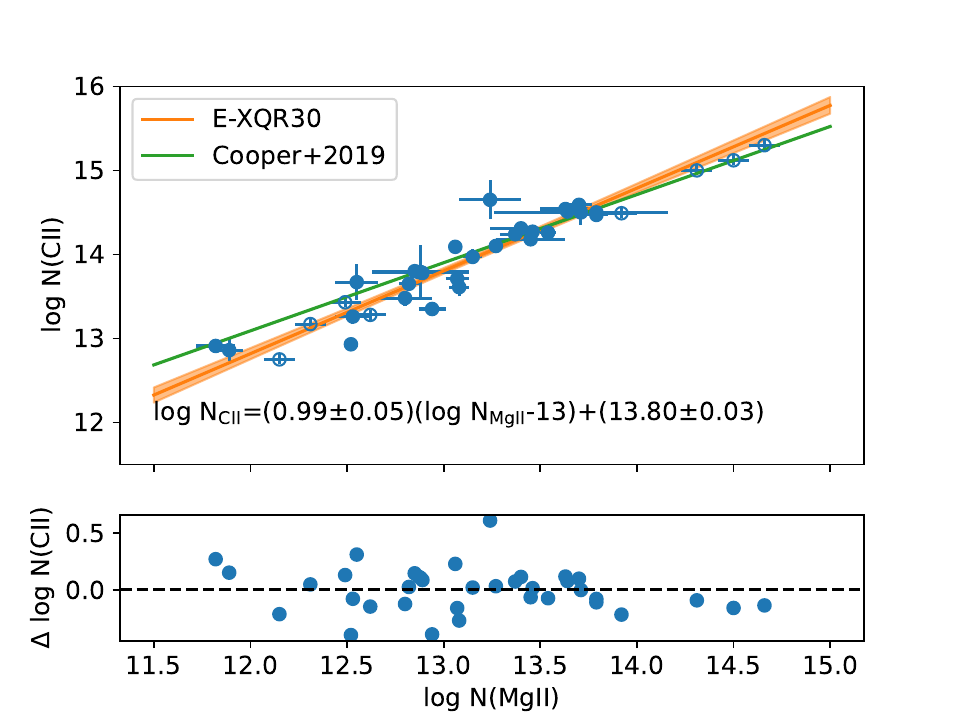}
    \caption{The column densities of \CII\ and \MgII\ for 29 absorption systems at 5.3~$< z <$~6.4 from the E-XQR-30 metal absorber catalogue. Measurements for systems with saturated absorption or evidence of partial covering are plotted as open markers. The orange line represents the linear fit to the E-XQR-30 data and the green line indicates the fit from \citet{Cooper_2019}. The bottom panel shows the residual scatter of the \CII\ column densities from the linear regression model.}
    \label{fig:almas plot}
\end{figure}

\subsection{Magnesium Measurements}
\CII~1334\AA\ cannot be detected at $z <$~5.17 in the E-XQR-30 sample due to overlap with the saturated Ly$\alpha$ forest. To quantify the N\textsubscript{CII}/N\textsubscript{CIV} ratio at lower redshifts, we use \MgII~2796\AA\ as a proxy for \CII. \MgII\ has an ionization potential of 7.6~eV, close to that of \CII, and \citet{Cooper_2019} demonstrated that there is a linear correlation between N\textsubscript{CII} and N\textsubscript{MgII}. We verify this correlation with the deeper E-XQR-30 dataset using the 37 systems where both \MgII\ and \CII\ are detected, spanning 5.3~$\leq z \leq$~6.4. We perform a linear fit using the orthogonal distance regression (ODR) method from the \verb|scipy.odr| \textsc{Python} package, accounting for the errors on the measurements. The \CII\ and \MgII\ column density measurements are shown in Figure~\ref{fig:almas plot}, with our linear fit plotted in orange and the best-fit from \citet{Cooper_2019} shown in green. Both correlations are consistent given the 1 sigma errors. Our best-fit relation between log(N\textsubscript{MgII}) and log(N\textsubscript{CII}) is given by the following equation: 
\begin{equation} 
\begin{split}
\rm log(N_{CII}) = (0.99 \pm 0.05) \left[ log(N_{MgII}) - 13 \right] + (13.80 \pm 0.03) \\
\end{split}
\label{eqn:MgII_CII}
\end{equation}
The slope is consistent with unity, suggesting that the ratio between the fraction of carbon in \CII\ and the fraction of magnesium in \MgII\ does not vary significantly with column density over log(N\textsubscript{MgII}/cm$^{-2}$)~=~12.0~--~14.0. The log(N\textsubscript{CII}/N\textsubscript{MgII}) ratio is constant at $\sim$0.8, very similar to the relative [C/Mg] abundance measured in the solar photosphere of 0.83 dex \citep{Asplund_2009}. We note that the 50\% completeness limit for detection of Mg~\textsc{ii} is log(N\textsubscript{MgII}/cm$^{-2}$)~$\simeq$~12.3, which based on Equation \ref{eqn:MgII_CII} corresponds to log(N\textsubscript{CII}/cm$^{-2}$)~$\simeq$~13.1; very close to the measured 50\% completeness limit for \CII\ of log(N\textsubscript{CII}/cm$^{-2}$)~$\simeq$~13.0 \citep{Davies_2023_Catalogue_paper}. This suggests that using \MgII\ as a proxy for \CII\ does not significantly change the column density sensitivity for detection of low-ionization absorbers in our sample.

We use Equation \ref{eqn:MgII_CII} to obtain \MgII\ column density measurements or upper limits for 426 systems. Of these, 61 have solid \MgII\ detections but no coverage of \CII, allowing us to use Equation 1 to infer N\textsubscript{CII}. We note that we find consistent results for the fraction of systems with low ionization, high ionization, and mixed absorption when using \CII\ or \MgII\ (see Section \ref{subsec:fractions}). 

\subsection{Column Density Limits}
\label{subsec:limit_calculations}
We calculate the 3$\sigma$ upper limit on the column density of each undetected line by summing the errors in quadrature over the relevant wavelength region to determine the maximum absorption equivalent width and then converting this equivalent width to a column density assuming it lies on the linear part of the curve of growth.

For each undetected feature, we consider that absorption may be present between $\lambda_{\rm obs}-\Delta \lambda/2$ and $\lambda_{\rm obs}+\Delta \lambda/2$. Within a single absorption system, different species can have different velocities and widths  (see \citealt{Davies_2023_Catalogue_paper}). To ensure that we cover all wavelengths where absorption could be present, we calculate $\Delta \lambda$ as the wavelength interval corresponding to a velocity range of 200~km/s. The vast majority of metal absorbers in E-XQR-30 have a 90\% velocity width (v$_{90}$) less than 200~km/s, and therefore this provides a conservatively high estimate for the equivalent width upper limit. The 1$\sigma$ upper limit on the equivalent width of undetected absorption is computed by summing the equivalent width error in quadrature over this wavelength range assuming uncorrelated noise:
\begin{equation}
\begin{split}
\rm dEW = \sqrt{\int_{\lambda_{\text{obs}} - \frac{\Delta \lambda}{2}}^{\lambda_{\text{obs}} + \frac{\Delta \lambda}{2}} \left(\frac{\text{error}(\lambda)}{\text{continuum}(\lambda)}\right)^2 d\lambda}  \\
\end{split}
\end{equation}

Here, d$\lambda$ is the width of each spectral channel in angstroms. The observed equivalent width limit is converted to rest frame by dividing by (1+$z$). To convert the rest frame EW limit to a column density limit, we use a modified version of the equation provided in section 4.2 of \citet{Cackett_2008}, which assumes that the absorption lies on the linear part of the curve of growth:
\begin{equation}
\rm dN (cm^{-2}) = {\frac{dEW_{rest}(\text{\AA})}{8.85 \times 10^{-21} \times \lambda_{rest}(\text{\AA})^2 \times \it{f}}}
\end{equation}
Here, $f$ denotes the oscillator strength. To obtain a 3$\sigma$ upper limit on the column density, we multiply the result obtained by 3. 

\subsection{Literature Comparison Samples}
We compare our E-XQR-30 measurements with the results of \citet{Cooper_2019}. They identified 69 systems with \MgII, \CIV\ and/or Fe~\textsc{ii} absorption at \mbox{5~$< z <$ 6.8} and then fit for \CII, \SiII, and \SiIV\ in the identified systems. This method, similar to that used by E-XQR-30, allows for the recovery of low ionization, high ionization, and mixed ionization systems. We use the data tables from \citet{Cooper_2019} to calculate the log(N\textsubscript{CII}/N\textsubscript{CIV}) and log(N\textsubscript{SiII}/N\textsubscript{SiIV}) ratios as well as the fraction of low ionization, high ionization, and mixed ionization systems. 

\begin{figure*}
 	\includegraphics[width=3.5in]{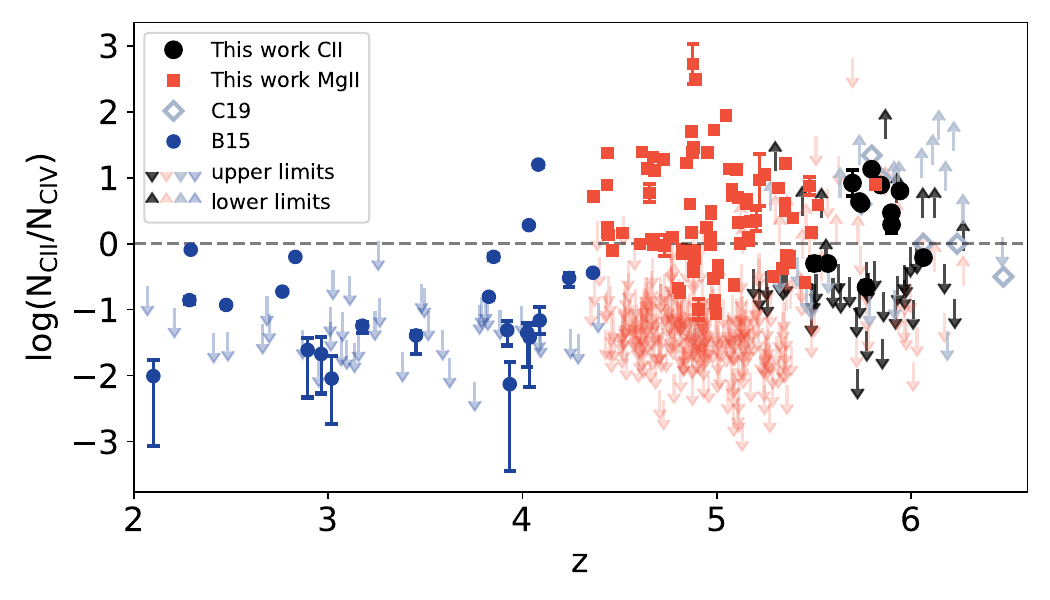}
        \includegraphics[width=3.5in]{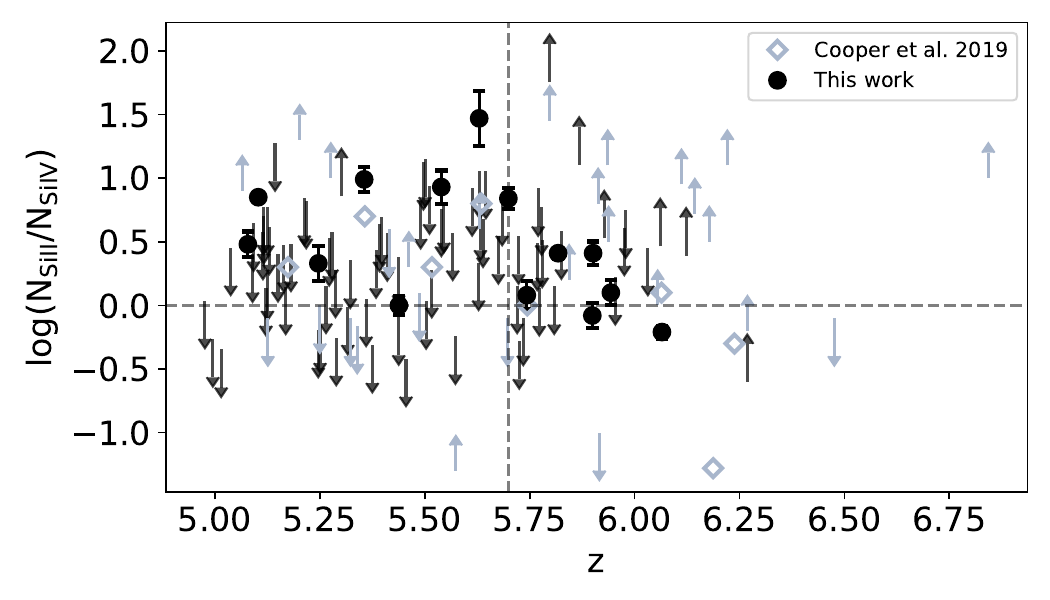}
    \vspace{-10pt}
    \caption{Evolution of log(N\textsubscript{CII}/N\textsubscript{CIV}) (left) and log(N\textsubscript{SiII}/N\textsubscript{SiIV}) (right) as a function of redshift with upper and lower limit values denoted by downward and upward arrows respectively. Our E-XQR-30 log(N\textsubscript{CII}/N\textsubscript{CIV}) and log(N\textsubscript{SiII}/N\textsubscript{SiIV}) measurements are shown in black, and measurements where \MgII\ is used as a proxy for \CII\ are shown in orange. The \citet{Cooper_2019} measurements (C19) are shown in grey and the \citet{Boksenberg_2015} (B15) measurements are shown in blue. The dashed horizontal lines mark the location for equal quantities of singly and triply ionized metals.}
    \label{fig:carbon_silicon_main_plot}
\end{figure*}

To extend our analysis to lower redshifts, we also utilise measurements from \citet{Boksenberg_2015} which characterises absorption systems at redshifts 2~$<z<$~4.4. Their paper presents comprehensive information on the column densities of several kinematic components in each absorption system, where systems are manually defined to encompass all velocity components within well-defined clumps of absorption with widths ranging up to a few hundred km~s$^{-1}$. For consistency with the E-XQR-30 and \citet{Cooper_2019} measurements, we sum the column densities of individual components to determine the total column density for each species in each absorption system. We only select systems for which all components have both N\textsubscript{CIV} and N\textsubscript{CII} measurements or limits and are not impacted by blending or other observational artifacts. When all components have robustly measured column densities, we directly sum the column densities and associated variance. If any component has a column density upper limit, that component is treated as having zero column density and the limit value is added (in quadrature) to the upper error on the column density of the system. This causes the column density measurements to have asymmetric errors. We note that only absorbers with \CIV\ detections were initially included in the \citet{Boksenberg_2015} sample. \CII-only systems are not considered, and therefore this sample cannot be used to compute the fraction of low-ionization, high-ionization and mixed absorbers at 2~$<z<$~4.4. For this reason, the \citet{Boksenberg_2015} sample is only used to examine how the log(N\textsubscript{CII}/N\textsubscript{CIV}) ratio evolves towards lower redshifts. When quantifying the line ratio evolution we account for differences between the column density sensitivities of the E-XQR-30 and \citet{Boksenberg_2015} datasets, as described in Section \ref{subsec:ratio_evolution}.

The properties of all datasets used in this analysis are summarized in Table~\ref{table:table_of_ions_and_systems}.

\section{Results}
\label{sec:results}
The left-hand panel of Figure~\ref{fig:carbon_silicon_main_plot} shows the evolution of log(N\textsubscript{CII}/N\textsubscript{CIV}) as a function redshift. Black points are E-XQR-30 measurements for systems where \CII\ is covered, red points are E-XQR-30 measurements using \MgII\ as a proxy for \CII, grey points show data from \citet{Cooper_2019} and blue points show data from \citet{Boksenberg_2015}. This figure does not display \MgII\ measurements for systems where \CII\ is accessible. Measurements for mixed systems are denoted by solid symbols with error bars, and arrows represent upper (lower) limits for systems with only high-ionization (low-ionization) absorption. The plot appears to reveal a decrease in log(N\textsubscript{CII}/N\textsubscript{CIV}) towards lower redshifts, primarily driven by the disappearance of systems with large log(N\textsubscript{CII}/N\textsubscript{CIV}) ratios at $z\lesssim$~4. We emphasize that the \citet{Boksenberg_2015} sample does not include \CII-only low-ionisation absorbers and therefore it is not possible to probe systems with lower limits on log(N\textsubscript{CII}/N\textsubscript{CIV}) at $z<$~4.3. The right-hand panel of Figure~\ref{fig:carbon_silicon_main_plot} shows the evolution of log(N\textsubscript{SiII}/N\textsubscript{SiIV}) over the redshift range covered by the E-XQR-30 and \citet{Cooper_2019} samples. There is no clear evidence for any variation in ionization properties with redshift, perhaps due to the much shorter redshift interval probed. 

We quantify the evolution in the ionization states of the metal absorbers using two methods. We first compute the fraction of low-ionization, high-ionization and mixed absorbers in 3 redshift bins (Section \ref{subsec:fractions}), and then perform linear regression to quantify the redshift evolution of the N\textsubscript{CII}/N\textsubscript{CIV} ratio for mixed absorbers where both species are detected (Section \ref{subsec:ratio_evolution}).

\begin{figure*}
        \includegraphics[width=3.2in]{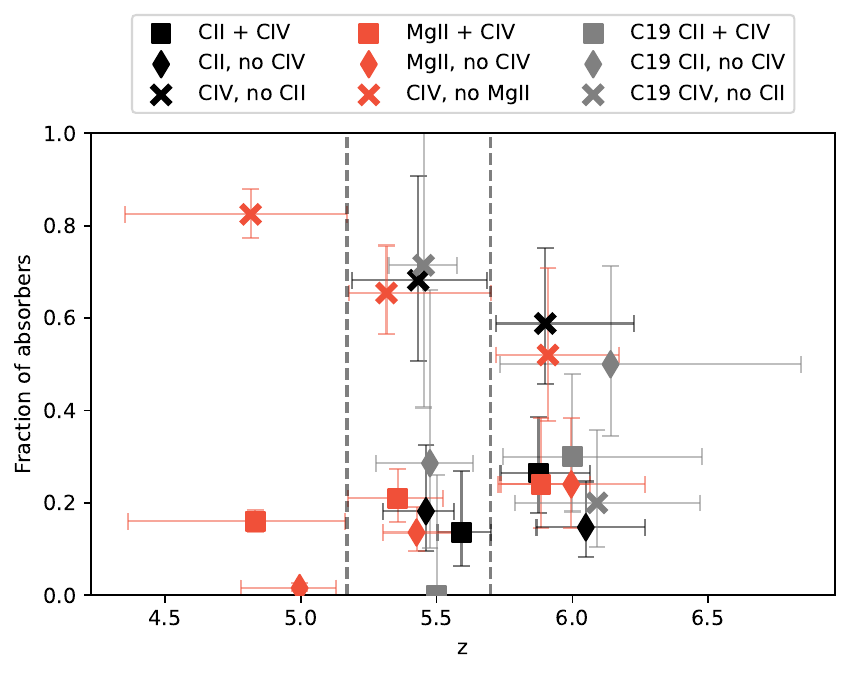}
	\includegraphics[width=3.2in]{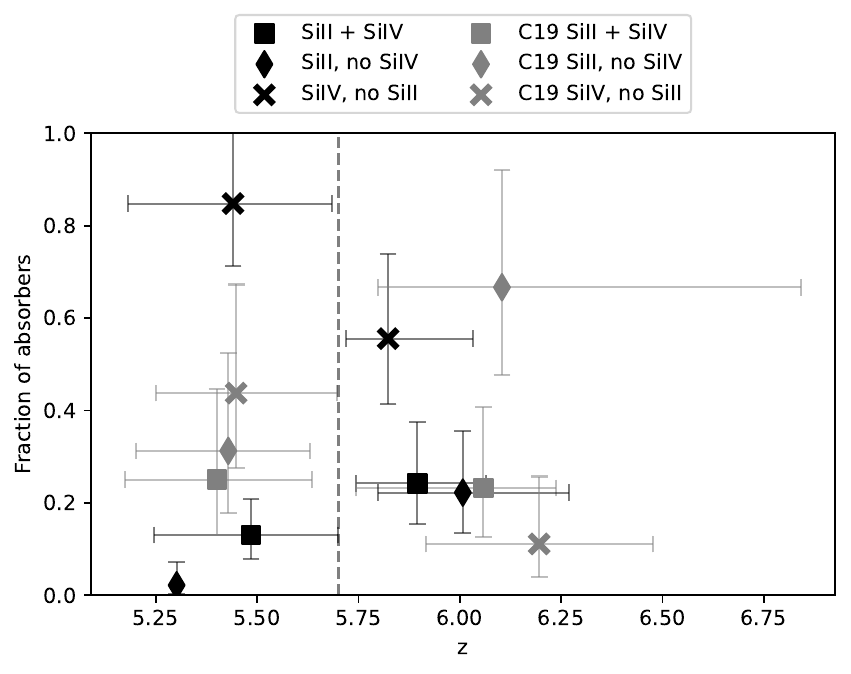}  
    \vspace{-10pt}
    \caption{The fraction of absorbers detected in carbon/magnesium (left) and silicon (right) with low-ionization, high-ionization or mixed absorption, split into redshift bins 4.34~$<z<$~5.17 (carbon only), 5.17~$<z<$~5.7 and 5.7~$<z<$~6.5. The color-coding of the samples is the same as in Figure \ref{fig:carbon_silicon_main_plot}. Horizontal error bars show the span of redshifts occupied by absorbers and vertical error bars show the 1$\sigma$ error on the fraction measurements. Some points have been offset by up to 0.02 on either axis for clarity.}
    \label{fig:fraction_plot}
\end{figure*}

\subsection{Fraction of Low-Ionization, High-Ionization and Mixed Absorbers}
\label{subsec:fractions}
For each combination of ions (\CII\ + \CIV, \MgII\ + \CIV, \SiII\ + \SiIV), we compute the number of absorption systems with only low ionization, only high ionization, or both ions detected, in 3 redshift bins: $z<$~5.17, 5.17~$<z<$~5.7 and $z>$~5.7 (Table~\ref{table:table_of_ions_and_systems}). The $z$~=~5.17 partition represents the lower boundary for \CII\ detection while $z$~=~5.7 is the redshift of the reported transition between low-ionization-dominated and high-ionization-dominated absorbers in \citet{Cooper_2019}. The \MgII-\CIV\ statistics include systems where both \CII\ and \MgII\ are accessible. We then compute the fraction of ions in each category as a function of redshift, shown in Figure~\ref{fig:fraction_plot}. The 1$\sigma$ errors on the fractions are computed assuming Poisson statistics, using Equations 9 and 12 from \citet{Gehrels_1986} with s~=~1 and $\beta$~=~0, appropriate for a 1$\sigma$ confidence interval. These equations are more accurate than the typical sqrt(N) approximation for small samples. 

The left-hand panel of Figure~\ref{fig:fraction_plot} reveals a strong evolution in the ionization state of carbon and magnesium absorbers over redshift. At 5.7~$< z <$~6.3, 52\% of absorbers have \CIV\ but no \MgII, 24\% have both \MgII\ and \CIV\ and 24\% have \MgII\ but no \CIV, whilst at \mbox{4.3~$<z<$~5.17}, 82\% of absorbers have \CIV\ but no \MgII, 16\% have both \MgII\ and \CIV\ and just 2\% have \MgII\ but no \CIV. Silicon absorbers, shown in the right panel, show similar trends to the carbon absorbers at $z>$~5.17 where both can be measured. The observed decrease in the incidence of low-ionization absorbers and corresponding increase in high-ionization absorbers towards lower redshifts is qualitatively consistent with previous findings \citep[e.g.][]{Becker_2019, DOdorico_2022, Davies_2023_CIV_paper, Sebastian_2024}. 

Interestingly, our observed evolution is much weaker than what was reported in \citet{Cooper_2019}. They found that 52\% of carbon absorption systems at $z >$~5.7 only show low ionization absorption; a factor of 2 larger than the 24\% reported in this study. This is likely driven by differences in the column density sensitivity of the datasets. The column density limit for the \citet{Cooper_2019} FIRE spectra is log(N/cm\textsuperscript{-2})~$\gtrsim$~13.5, compared to 50\% completeness limits of log(N\textsubscript{CIV}/cm\textsuperscript{-2})~=~13.2 and log(N\textsubscript{CII}/cm\textsuperscript{-2})~=~13.0 for E-XQR-30. 81\% of the \CIV\ absorbers detected by E-XQR-30 lie below the 50\% completeness limit in \citet{Cooper_2019}, and the 2~--~3 times deeper E-XQR-30 observations show that \CIV\ absorbers heavily dominate the absorber population by number density at all redshifts (Figure \ref{fig:fraction_plot} left). To investigate the impact of sensitivity on the measured absorber fractions, we re-calculated the absorber fractions only including detections above the detection limit of the FIRE spectra. \citet{Cooper_2019} quote an approximate detection limit of log(N\textsubscript{CII}/cm\textsuperscript{-2})~$>$~13.5, which based on Equation \ref{eqn:MgII_CII} corresponds to log(N\textsubscript{MgII}/cm\textsuperscript{-2})~$\gtrsim$~12.7. Based on results from E-XQR-30 \citep{Davies_2023_Catalogue_paper}, we assume a comparable detection limit in CIV, i.e. log(N\textsubscript{CIV}/cm\textsuperscript{-2})~$>$~13.5. We found that imposing this cut increases the fraction of \MgII-only systems at $z>5.7$ by more than a factor of 2, from 24\% to 56\% such that these absorbers dominate the population at that redshift, consistent with the \citet{Cooper_2019} results. The fraction of \MgII-only systems at \mbox{5.17 $<z <$ 5.7} also increases from 2\% to 25\%, but \CIV-only absorbers are still dominant, accounting for 45\% of systems at this redshift. These results suggest that the balance between high-ionization and low-ionization gas may differ for strong and weak absorbers.

Nevertheless, our results support the existence of a significant population of low-ionization-only absorbers at $z>$~5.7. Such systems are not expected based on our canonical picture of the low redshift CGM whereby dense, low-ionization clouds with a small volume filling factor are surrounded by higher ionization clouds with a larger covering fraction. In this scenario, we would expect metal absorber samples to be comprised primarily of high-ionization and mixed absorbers, with very few systems displaying only low-ionization absorption.

We further explore the characteristics of the different absorber populations by comparing their kinematics quantified using v$_{90}$ which is the velocity interval enclosing 90 per cent of the optical depth for a single transition of a given absorption species. Combining all redshift ranges, \CIV-only absorbers have an average v$_{90}$ of \mbox{98~$\pm$~56 km s$^{-1}$} while \CII-only and \MgII-only absorbers have a lower average v$_{90}$ of \mbox{68~$\pm$~27 km s$^{-1}$}. This is consistent with the overall statistics from the E-XQR-30 metal absorber catalog which show that individual components of high ionization absorbers generally have larger linewidths than components of low ionization absorbers \citep{Davies_2023_Catalogue_paper}. Our findings are also consistent with many previous works suggesting that high ionization absorbers on average are expected to arise from warmer and more turbulent material \citep[e.g.][]{Wolfe_1993, Rauch_1996, Lehner_2014, Muzahid_2014, Fox_2015, Pradeep_2020}. Interestingly, mixed ionization systems have the highest average v$_{90}$ of \mbox{157~$\pm$~79 km s$^{-1}$} for \CIV\ absorption and \mbox{108~$\pm$~67 km s$^{-1}$} for \MgII\ absorption, suggesting they trace more complex multiphase systems, consistent with the findings of \citet{Codoreanu_2018}. There is weak evidence for an increase in the average v$_{90}$(\CIV) of mixed absorbers towards lower redshift, from \mbox{125 $\pm$ 71 km s$^{-1}$} at $z > 5.17$ to \mbox{175 $\pm$ 80 km s$^{-1}$} at \mbox{4.3 $<z <$ 5.17}. This is primarily driven by a slight increase in the average number of kinematic components per absorption system from 1.3 to 1.7, with no significant change in the turbulent $b$ parameter of individual components.

\citet{Cooper_2019} suggest that the detection of a significant population of \CII-only absorbers at $z>$~5.7 may indicate that the CGM at this redshift has a lower ionization state (see also \citealt{Becker_2019} and/or is significantly more metal-poor than the CGM at $z\sim$~3. These possibilities will be discussed further in Section 4. 

\subsection{The log(N\textsubscript{CII}/N\textsubscript{CIV}) Column Density Ratio}
\label{subsec:ratio_evolution}
\subsubsection{Redshift Evolution}
\begin{figure*}
	\includegraphics[width=5.3in, clip = True, trim = 0 10 0 0]{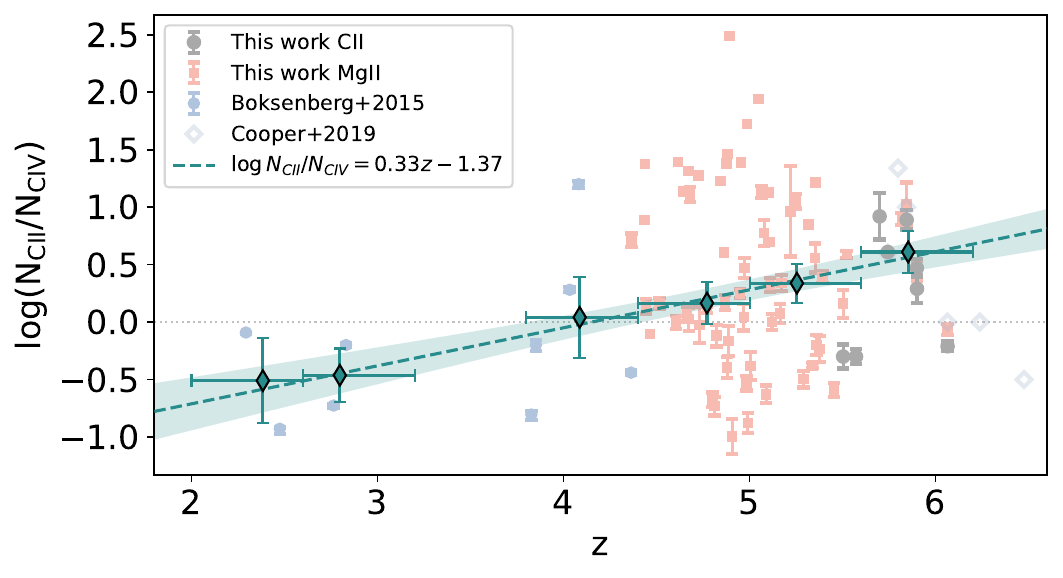}
    \caption{Measurements of the log(N\textsubscript{CII}/N\textsubscript{CIV}) ratio for mixed absorption systems with column densities above the 50\% completeness limit of the E-XQR-30 catalog (log(N\textsubscript{CII})~$>$~13.0, log(N\textsubscript{CIV})~$>$~13.2 (background)). Green solid markers show medians and associated error (1.253$\sigma$/$\sqrt{N}$) in bins of redshift. The linear fit to these medians reveals a clear decline in log(N\textsubscript{CII}/N\textsubscript{CIV}) towards lower redshift.}
    \label{fig:trendline_fig}
\end{figure*}

We quantify the evolution of log(N\textsubscript{CII}/N\textsubscript{CIV}) as a function of redshift by fitting a linear relation. We only use mixed systems when calculating the redshift evolution of the N\textsubscript{CII}/N\textsubscript{CIV} ratio because the low and high ionization systems likely trace predominately single-phase gas and therefore the lower and upper limits on the N\textsubscript{CII}/N\textsubscript{CIV} ratios measured for these systems mostly reflect the measurement uncertainties rather than the intrinsic properties of the absorption systems. In addition, we only consider systems with column densities above the 50\% completeness limit of the E-XQR-30 sample (log(N\textsubscript{CII})~$>$~13.0, log(N\textsubscript{CIV})~$>$~13.2) to prevent biases that may arise due to comparing systems with different column densities at different redshifts. The background markers in Figure~\ref{fig:trendline_fig} show the measurements included in this analysis.

The vast majority of our log(N\textsubscript{CII}/N\textsubscript{CIV}) measurements are at \mbox{4.3~$\lesssim z \lesssim$~5.17}, so to prevent the best-fit being solely driven by this redshift interval, we bin the data and compute the median log(N\textsubscript{CII}/N\textsubscript{CIV}) ratio and the associated error (1.253$\sigma$/$\sqrt{N}$) in each redshift bin. We try bin sizes of \mbox{0.2 $< \Delta(z) <$ 1.0} and select a bin size of $\Delta(z)$~=~0.6 which produces the best compromise between number of bins and number of objects per bin. Our median measurements are shown by the green markers and the best fit line is over-plotted in green with the shaded region indicating the $\pm$~1$\sigma$ error interval.

For the fiducial bin size of $\Delta(z)$~=~0.6, we measure a slope of 0.33~$\pm$~0.08. Our results indicate that the N\textsubscript{CII}/N\textsubscript{CIV} ratio declines by a factor of $\sim$20 between $z\sim$~6 and $z\sim$~2, with a significance of $\sim$~4$\sigma$. We test the impact of the chosen bin size on this result and find that for bin sizes of 0.2 $< \Delta(z) <$ 1.0, the slope varies from 0.26~--~0.34, within the margin of error of the original estimate. There is some evidence that the N\textsubscript{CII}/N\textsubscript{CIV} ratio decreases slightly over the E-XQR-30 redshift range from $z\sim$~6.3 to $z\sim$~4.3, primarily because most of mixed absorbers at $z\gtrsim$~5.5 are dominated by high ionization absorption with log(N\textsubscript{CII}/N\textsubscript{CIV})~$>$~0. However, this evolution is not statistically significant due to the small number of systems in the highest redshift bin and the large intrinsic scatter in the $z\simeq$~5 measurements.

The decrease in the average N\textsubscript{CII}/N\textsubscript{CIV} ratio of mixed absorbers towards lower redshift supports the conclusions from the fraction analysis that the average ionization state of metal absorbers increases over cosmic time.

\subsubsection{Correlation with ion column densities}
\begin{figure*}
	\includegraphics[scale=0.55, clip = True, trim = 10 10 0 0]{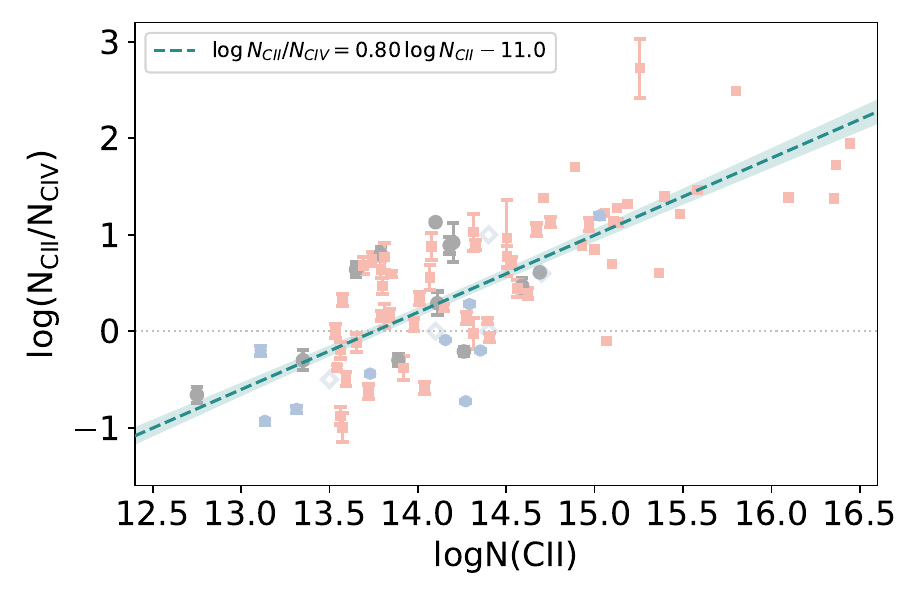}
    \includegraphics[scale=0.55, clip = True, trim = 10 10 0 0]{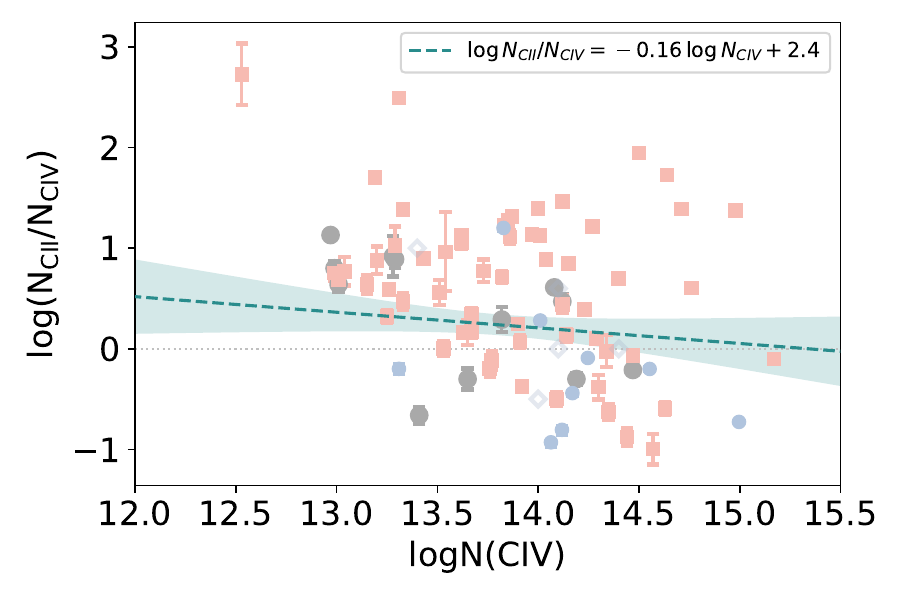} \\
    \includegraphics[scale=0.55, clip = True, trim = 10 10 0 0]{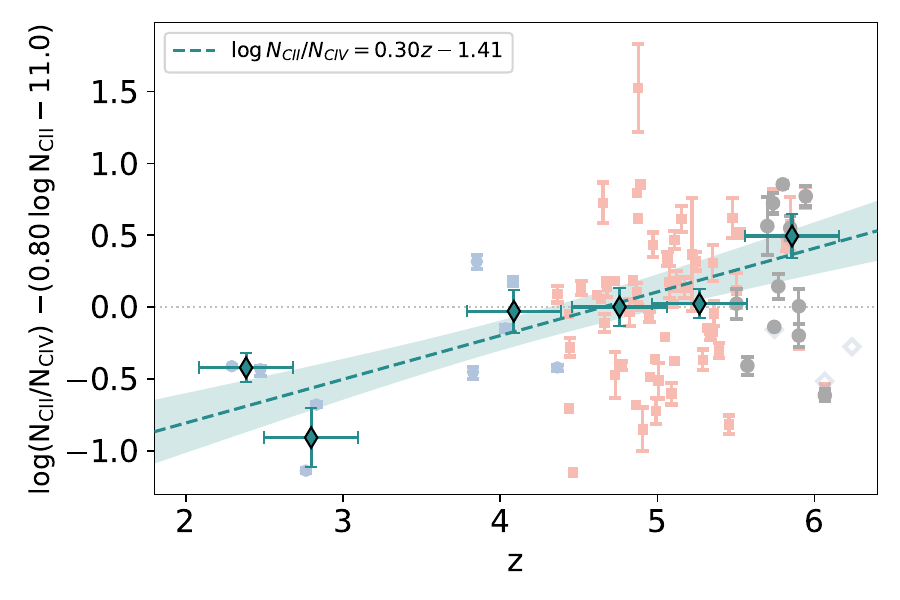}
    \includegraphics[scale=0.55, clip = True, trim = 10 10 0 0]{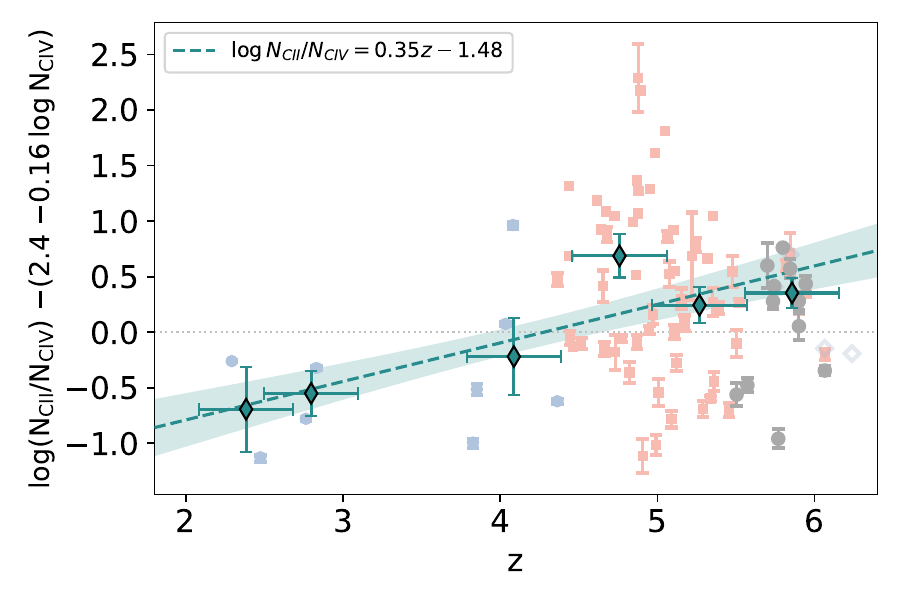}
    \caption{Top: log(N\textsubscript{CII}/N\textsubscript{CIV}) correlates strongly with log(N\textsubscript{CII}) (left) but shows little dependence on log(N\textsubscript{CIV}) (right). Bottom: accounting for the correlations with log(N\textsubscript{CII} (left) or log(N\textsubscript{CIV} (right) reduces the scatter in log(N\textsubscript{CII}/N\textsubscript{CIV}) at fixed redshift but does not significantly impact the observed redshift evolution. Plot symbols are the same as in Figure \ref{fig:trendline_fig}.}
    \label{fig:ratio_dependencies}
\end{figure*}

Figure \ref{fig:ratio_dependencies} examines how the log(N\textsubscript{CII}/N\textsubscript{CIV}) ratio correlates with the column densities of the individual ions: log(N\textsubscript{CII}) (top left) and log(N\textsubscript{CIV}) (top right). The results are striking: the column density ratio shows a strong positive correlation with log(N\textsubscript{CII}) but is virtually independent of log(N\textsubscript{CIV}).

To investigate whether the redshift evolution in log(N\textsubscript{CII}/N\textsubscript{CIV}) is driven by a decrease in the typical log(N\textsubscript{CII}) towards lower redshifts, we calculated the expected log(N\textsubscript{CII}/N\textsubscript{CIV}) of each absorber based on its measured log(N\textsubscript{CII}), subtracted this from the measured ratio, and then plotted the residuals as a function of redshift. The results are shown in the bottom left panel of Figure \ref{fig:ratio_dependencies}. Accounting for the trend with log(N\textsubscript{CII}) significantly reduces the scatter at fixed redshift compared to Figure \ref{fig:trendline_fig} but does not significantly change the slope of the redshift evolution. If we instead compute the residuals from the log(N\textsubscript{CIV}) correlation, the ratios remain essentially unchanged (see bottom right panel of Figure \ref{fig:ratio_dependencies}), as expected given that the log(N\textsubscript{CIV}) trend is relatively weak.

Our results indicate that the redshift evolution in the log(N\textsubscript{CII}/N\textsubscript{CIV}) ratio is not driven by differences in the typical column densities of absorbers probed at each redshift. The top panels show that we probe similar column densities at all redshifts, which is a direct consequence of the fact that we remove systems from the \citet{Boksenberg_2015} sample that fall below the detection limit of E-XQR-30. Furthermore, the redshift trends remain essentially unchanged even after accounting for correlations with column density.

\section{Discussion}
\label{sec:modelling}

The unprecedented depth and sample size of the E-XQR-30 catalog have enabled us to robustly demonstrate that the ionization state of circumgalactic gas evolves significantly in the early Universe. The fraction of low-ionization absorbers declines significantly over a period of less than 400~Myr, from 24\% at $z\sim$~6 to just 2\% at $z\sim$~4.7. Furthermore, the typical N\textsubscript{CII}/N\textsubscript{CIV} ratio of mixed absorbers declines by a factor of 20 over the 2~Gyr from $z\sim$~5.9 to $z\sim$~2.3. 

We used the \textsc{Cloudy} photoionization code \citep{Ferland_2017} to investigate the physical mechanisms driving the decline in the N\textsubscript{CII}/N\textsubscript{CIV} ratio towards lower redshifts. We run models of CGM clouds with solar-scaled abundances \citep{Grevesse2010} illuminated using the \citet[][hereafter HM12]{Haardt_2012} UV background model. Following the approach of \citet{Cooper_2019}, we explore the impact of varying hydrogen volume density and metallicity at fixed redshift ($z$~=~4, 5, 6) and N\textsubscript{HI} (10\textsuperscript{15}~cm\textsuperscript{-2}, 10\textsuperscript{17}~cm\textsuperscript{-2} and 10\textsuperscript{19}~cm\textsuperscript{-2}), covering column densities representative of environments ranging from the densest regions of the intergalactic medium to the neutral circumgalactic medium \citep[e.g.][]{Peroux_2020}. We note that fixing the \ion{H}{I} column density makes the results relatively insensitive to the choice of UVB model: in Appendix \ref{appendix:UVBs} we show that consistent results are obtained using the \citet{Puchwein_2019} and \citet{Faucher-Giguere_2020} UV background models which are in better agreement with estimates of the \ion{H}{i} photoionization rate at z=5$-$6. We caution though that these photoionization models are quite simple, as \CII\ and \CIV\ absorbers likely trace gas at different column densities and metallicities.

\begin{figure*}
        \includegraphics[width=7in]{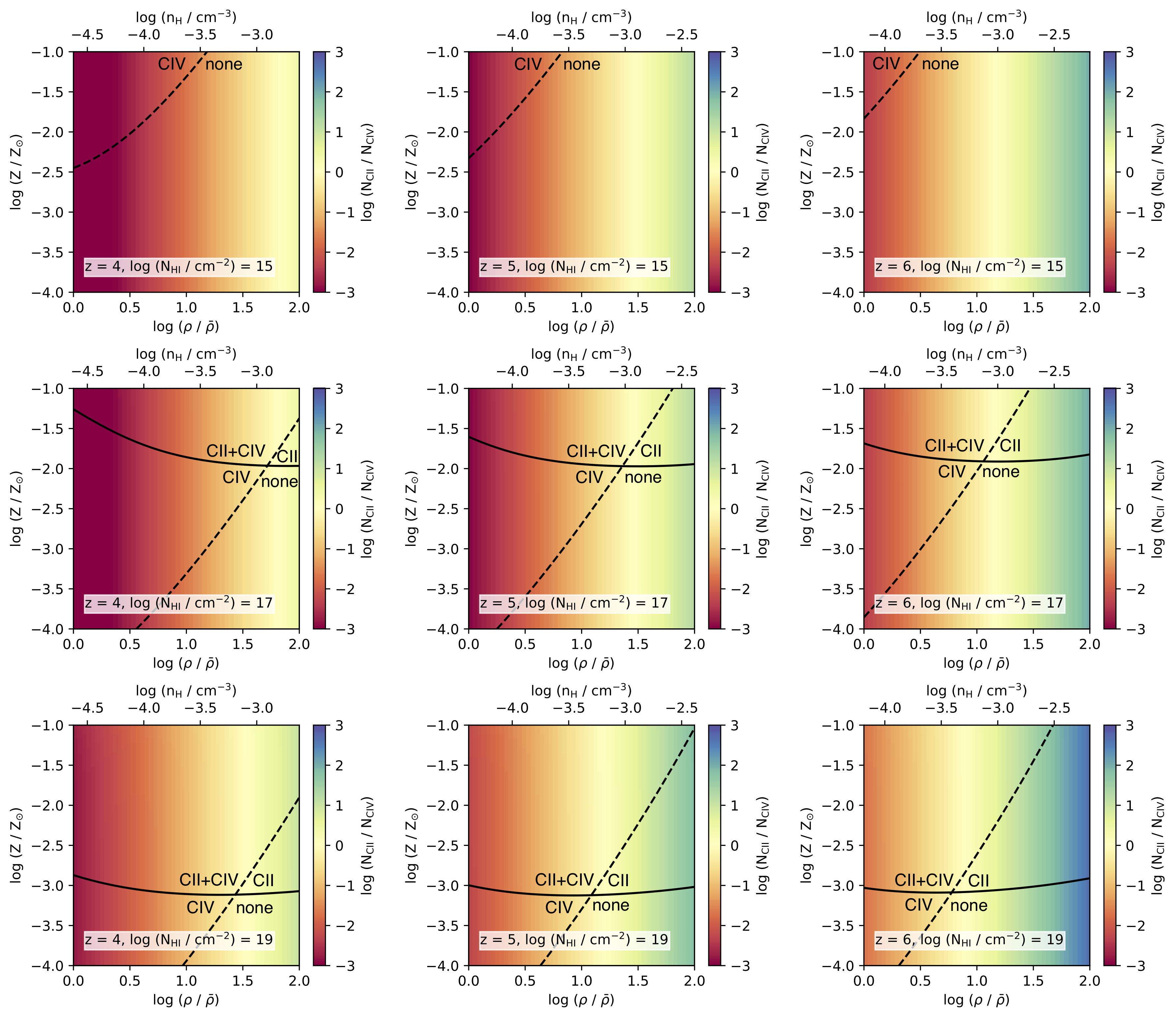} 
    \vspace{-5pt}
    \caption{Photoionization models of log(N\textsubscript{CII}/N\textsubscript{CIV}) as a function of metallicity ($y$-axis), hydrogen column density (upper $x$-axis), and density normalized to the cosmic mean at the relevant redshift (lower $x$-axis) for gas clouds at fixed neutral hydrogen column density (rows) and redshift (columns), illuminated by a \citet{Haardt_2012} UV background. The range of density in units of the cosmic mean is the same in all panels, while the corresponding range of proper hydrogen number density changes with redshift. The black solid (dashed) lines mark the column density of the 50\% completeness limit for \CII\ (\CIV) absorbers in the E-XQR-30 dataset, corresponding to N\textsubscript{CII} = 10\textsuperscript{13}~cm\textsuperscript{-2} and N\textsubscript{CIV} = 10\textsuperscript{13.2}~cm\textsuperscript{-2}. These different quadrants indicate the detectability of \CII\ and \CIV\ absorption: \CII-only, no detection, \CIV-only, and \CII~+~\CIV. For the lowest \ion{H}{i} column density considered here, N\textsubscript{HI}  = 10\textsuperscript{15}~cm\textsuperscript{-2}, \ion{C}{ii} would not be detectable for gas in the metallicity and density range probed by these models.}
    \label{fig:modelling}
\end{figure*}

The photoionization model results are shown in Figure~\ref{fig:modelling}. The black solid (dashed) lines delineate the 50\% completeness limit for detection of \CII\ (N\textsubscript{CII} = 10\textsuperscript{13}~cm\textsuperscript{-2}) and \CIV\ (N\textsubscript{CIV} = 10\textsuperscript{13.2}~cm\textsuperscript{-2}) in the E-XQR-30 dataset. The column densities of both species correlate strongly with metallicity, whilst the \CIV\ column density also decreases towards higher hydrogen volume density where carbon is primarily found in lower ionization states (see also \citealt{Davies_2023_Catalogue_paper}). In contrast, the \CII\ column density is relatively independent of hydrogen volume density, as it is more sensitive to the ionization parameter. Whether a \CII-only absorber, \CIV-only absorber,  \CII~+~\CIV\ absorber or no detection would occur at a given hydrogen volume density \ion{H}{i} column density depends on the metallicity and probed by the absorbers. It is likely that not all of the parameter space shown here may not correspond to realistic properties of true $z$~=~4$~-~$6 absorbers, e.g., it is not expected to see \CIV-only systems with \mbox{N\textsubscript{HI} = 10\textsuperscript{19}~cm\textsuperscript{-2}}. These models do, however, demonstrate that a wide range of values for N\textsubscript{CII}/N\textsubscript{CIV} are possible, depending on the physical properties of the absorbers. This is in good agreement with the spread of values measured in Figure~\ref{fig:carbon_silicon_main_plot}, especially when considering the upper and lower limits.

\subsection{Impact of metallicity}

The first important result demonstrated by Figure~\ref{fig:modelling} is that our ability to detect \CII\ and \CIV\ is strongly metallicity-dependent. Focusing on the upper middle panel, at fixed column density ratio \mbox{log(N\textsubscript{CII}/N\textsubscript{CIV}) = $-1$}, an absorber might be detected in both \CII\ and \CIV, or \CIV\ only, or neither depending on its metallicity. The impact of metallicity on the detectability of \CII\ and \CIV\ is especially relevant at high redshifts because the gas giving rise to these lines is likely not cospatial and the metallicity may differ significantly between \CII-hosting and \CIV-hosting regions \citep[e.g.][]{Cooper_2019}. The denser \CII-hosting regions are likely to reside closer to galaxies and therefore become enriched with metals earlier than the more diffuse \CIV-hosting regions further out in the halo. Star-forming galaxies during the epoch of reionization are embedded in massive reservoirs of \CII-emitting cold gas \citep[e.g.][]{Fujimoto_2019, Ginolfi_2020, Bischetti_2024, Bischetti_2025} which simulations predict are likely enriched by feedback processes \citep[e.g.][]{Vito_2022, Pizzati_2023}. \citet{Davies_2023_CIV_paper} showed that the rapid evolution in the \CIV\ cosmic mass density between $z\sim$~6.3 and $z\sim$~4.3 could be explained by chemical enrichment if the delay time for outflows enrich diffuse gas in the outer halo is on the order of \mbox{100~--~400~Myr}. This is also consistent with the measured outflow velocities and impact parameters of \CIV\ absorbers at this redshift. The upper right panel of Figure~\ref{fig:modelling} suggests that, at $z\sim$~6, diffuse, low metallicity gas in the halo could remain mostly undetected, whilst denser, more enriched gas could be detected as \CII-only or \CII~+~\CIV\ absorption. At $z\sim$~4, the predicted detection statistics are significantly different, favouring \CII+\CIV\ and \CIV-only absorption and disfavouring \CII-only absorption. This predicted shift is consistent with the ionization fraction trends observed in Figure~\ref{fig:fraction_plot}. The diversity in observed \CII-only, \CIV-only and mixed systems at fixed redshift could be explained if the E-XQR-30 absorbers show a spread in metallicity as wide as what is observed at $z\sim$~3 \citep[e.g.][]{Lehner_2022} and there is also some variation in the gas density and N\textsubscript{HI} probed by the absorbers. 

\subsection{Impact of H~\textsc{i} column density}

Comparing the grids in the top, middle and bottom rows of Figure~\ref{fig:modelling}, it is clear that the observed absorber statistics are strongly dependent on H~\textsc{i} column density. For N\textsubscript{H~\textsc{i}}~=~10$^{19}$~cm$^{-2}$, \CII-only and \CII~+~\CIV\ absorbers are expected to dominate at most metallicities, whilst for N\textsubscript{H~\textsc{i}}~=~10$^{17}$~cm$^{-2}$, \CIV-only absorbers become much more prominent at low metallicities and densities. At N\textsubscript{H~\textsc{i}}~=~10$^{15}$~cm$^{-2}$, only \CIV~ systems are expected to be detectable. If the typical H~\textsc{i} column densities of metal absorption systems decrease towards lower redshifts (as the UV background evolves), the fraction of low-ionization, high-ionization and mixed absorbers would change in a qualitatively similar manner to what is observed in E-XQR-30.

Islands of neutral gas act as effective sinks for ionizing photons, and as these neutral islands disappear, the mean free path of ionizing photons increases significantly \citep{Becker_2021, Gaikwad2023, zhu_2023, Satyavolu24}. This is expected to drive a rapid increase in the strength of the UV background \citep{lewis_2022} between $z\sim$~6 and $z\sim$~5, consistent with observations showing that the \ion{H}{i} photoionization rate rises sharply over this redshift range \citep{Calverley_2011, Gaikwad2023, Davies_2024}. 

The increase in the \ion{H}{i} photoionization rate directly impacts the density at which gas begins to self shield and is therefore expected to induce rapid evolution in the typical column density of H~\textsc{i} absorbers \citep{Schaye_2001}. Assuming the evolution in \ion{H}{i} photoionization rate measured by \citet{Gaikwad2023}, the threshold self-shielding density would increase by a factor of approximately 2.5 between $z\sim$~6 and $z\sim$~5, meaning that the circumgalactic medium will be self-shielded to smaller and smaller radii as the amplitude of the UVB increases \citep{Sadoun_2017}.

The shape of the H~\textsc{i} column density distribution function is difficult to measure directly near the EoR due to the saturation of the Ly$\alpha$ forest \citep{Fan_2006}. The highest redshift measurement to date is at $z\sim$~5 \citep{Crighton_2019}, which still shows a smooth evolution when compared to lower redshift data. However, there is some evidence that the H~\textsc{i} column density distribution should evolve more rapidly towards higher redshifts. The incidence of Lyman-limit systems (N\textsubscript{H~\textsc{i}}~=~10$^{17.2}$~--~10$^{19}$~cm$^{-2}$) is observed to increase with redshift towards $z\sim$~6 \citep{Songaila_2010}, whilst the incidence of DLAs (N\textsubscript{H~\textsc{i}}~$>$~10$^{20.3}$~cm$^{-2}$) appears to peak at $z\sim$~4 and decrease significantly towards higher redshifts \citep{Oyarzun_2025}. The fraction of galaxies showing strong Ly$\alpha$ emission decreases towards high redshift, with some evolution already seen between redshifts 5 and 6 \citep{Tang_2024}. If the typical H~\textsc{i} column density of a metal absorber host evolves from something like a Lyman-limit system at lower redshifts, to something closer to a sub-DLA \mbox{(N\textsubscript{H~\textsc{i}}~=~10$^{19}$~--~10$^{20.3}$~cm$^{-2}$)} at higher redshifts, this may explain the change in the relative fractions of low-ionization, high-ionization and mixed absorbers with redshift. In this scenario, the metal absorbers would be in good agreement with other indirect probes of the evolution of H~\textsc{i} absorbers in ionized regions towards the end of reionization. 

The majority of the scatter in the observed log(N\textsubscript{CII}/N\textsubscript{CIV}) ratios can likely be explained by local fluctuations in the gas density, ionization parameter and metallicity. Prior to the end of reionization, spatial fluctuations in the UV background cause large-scale variations in the mean flux of the Ly$\alpha$ forest \citep{Bosman_2022} and may also enhance the scatter in \mbox{log(N\textsubscript{CII}/N\textsubscript{CIV})} above $z\simeq$~5. However, there are too few measurements at $z<$~4.5 to enable a robust measurement of intrinsic scatter as a function of redshift.

\section{Conclusions}
\label{sec:conclusions}
We have used a sample of 488 metal absorption systems at \mbox{4.3~$\lesssim z \lesssim$ 6.3} from the E-XQR-30 metal absorber catalog combined with 75 absorption systems at \mbox{2~$\lesssim z \lesssim$ 4} from \citet{Boksenberg_2015} to investigate the evolution in the ionization properties of circumgalactic gas from the epoch of reionization to cosmic noon. Our results are based on analysis of the singly and triply ionized ions of carbon (\CII\ and \CIV) and silicon (\SiII\ and \SiIV). We also use \MgII\ as a proxy for \CII\ to enable us to extend the analysis to $z<$~5.17 where \CII\ falls in the Ly$\alpha$ forest.

Using 3 pairs of ions (\CII~+~\CIV, \MgII~+~\CIV, \SiII~+~\SiIV), we classify absorption systems as low-ionization only, high-ionization only, or mixed ionization and investigate how the fraction of absorbers in each category evolves with redshift. For mixed absorbers where both ions are robustly detected, we compute the column density ratios N\textsubscript{CII}/N\textsubscript{CIV} and N\textsubscript{SiII}/N\textsubscript{SiIV} and investigate how these evolve over redshift. Our main conclusions are as follows:

\begin{enumerate}
    \item High-ionization systems dominate the absorber population by number density even at $z\sim$~6, contrary to previous findings. We recover a significant population of weak \CIV\ absorbers that were likely undetected in previous large samples with lower spectral resolution and signal-to-noise.

    \item The fraction of low-ionization systems (with detected \MgII\ but no \CIV) declines significantly over a period of less than 400 Myr, from 24\% at $z\sim$~6 to just 2\% at $z\sim$~4.7. The fraction of high-ionization systems (with \CIV\ but no \MgII) increases correspondingly from 52\% at $z\sim$~6 to 82\% at $z\sim$~4.3, with little change in the fraction of mixed systems.

    \item The average N\textsubscript{CII}/N\textsubscript{CIV} ratio of mixed absorbers declines by a factor of $\sim$~20 from $z\sim$~6 to $z\sim$~2, providing additional evidence for an increase in the ionization state of circumgalactic gas from the epoch of reionization to cosmic noon. 

    \item By running large grids of \textsc{Cloudy} models, we show that at fixed n\textsubscript{H} and N\textsubscript{HI}, the detectability of \CII\ and \CIV\ is strongly dependent on metallicity. This is especially relevant at high redshifts because the denser \CII-hosting regions reside closer to galaxies and likely become enriched with metals significantly earlier than the more diffuse \CIV-hosting regions further out in the halo. The models suggest that at $z\sim$~6, diffuse, low metallicity halos gas could be detected at low \CIV\ column density, whilst denser, more enriched gas could be detected as \CII-only or \CII~+~\CIV\ absorption. At $z\sim$~4, this would transition to \CII~+~\CIV\ absorption closer to galaxies and stronger \CIV-only absorption at larger distances. The observed diversity in \CII-only, \CIV-only and \CII~+~\CIV\ absorbers at fixed redshift can likely be explained by spreads in metallicity, gas volume density and H~\textsc{i} column density.
    
    \item A decrease in the typical H~\textsc{i} column density of metal absorbers towards lower redshifts, linked to the declining average cosmic mean density, would also lead to increasing dominance of high-ionization absorbers at lower redshifts. We hypothesize that the rapid evolution in the ionization states of metal absorbers between $z\sim$~6 and $z\sim$~5 could be driven by rapid changes in their H~\textsc{i} column densities resulting from the disappearance of neutral islands and a corresponding increase in the mean free path of ionizing photons during the final stages of reionization. The timing of this transition is consistent with recent findings that reionization ends at $z$=5.44$\pm$0.02 \citep{Qin_2024}.
\end{enumerate}

Our results confirm that the ionization states of metal absorbers evolve significantly in the early Universe and highlight the complexities of disentangling the various factors contributing to this evolution.

\section*{Acknowledgements}
We thank the referee for their thoughtful comments which improved the clarity of the manuscript. SR thanks Chris Blake for helpful discussions regarding the statistical analysis in this paper and acknowledges the Wurundjeri people of the Kulin nation who are the traditional custodians of the lands on which this research was conducted. This research was supported by the Australian Research Council Centre of Excellence for All Sky Astrophysics in 3 Dimensions (ASTRO 3D), through project number CE170100013. RLD is supported by the Australian Research Council through the Discovery Early Career Researcher Award (DECRA) Fellowship DE240100136 funded by the Australian Government. SEIB is supported by the Deutsche Forschungsgemeinschaft (DFG) under Emmy Noether grant number BO 5771/1-1. HC thanks the support by the Natural Sciences and Engineering Research Council of Canada (NSERC), funding reference \#RGPIN-2025-04798 and \#DGECR-2025-00136, and by the University of Alberta, Augustana Campus. VD and LW acknowledge financial support from the Bando Ricerca Fondamentale INAF 2022 Large Grant ``XQR-30''. MH and GK acknowledge support by UKRI-STFC (grant reference ST/Y004191/1). GK is supported by the Department of Atomic Energy (Government of India) research project with Project Identification Number RTI 4002. Based on observations collected at the European Organisation for Astronomical Research in the Southern Hemisphere under ESO Programme IDs 0100.A-0625, 0101.B-0272, 0102.A-0154, 0102.A-0478, 084.A-0360(A), 084.A-0390(A), 084.A-0550(A), 085.A-0299(A), 086.A-0162(A), 086.A-0574(A), 087.A-0607(A), 088.A-0897(A), 091.C-0934(B), 096.A-0095(A), 096.A-0418(A), 097.B-1070(A), 098.B-0537, 098.B-0537(A), 1103.A-0817, 294.A-5031(B), 60.A-9024(A). 

\section*{Data Availability}
The metal absorber catalogue used in this paper is publicly available
and can be downloaded from this GitHub repository: \href{https://github.com/XQR-30/Metal-catalogue}{https://github.com/XQR-30/Metal-catalogue}.



\bibliographystyle{mnras}
\bibliography{ref} 




\appendix

\section{Impact of UV background model on properties of detectable ions}
\label{appendix:UVBs}

Throughout this work, we have explored the evolution of the N\textsubscript{CII}/N\textsubscript{CIV} ratio using the \citet{Haardt_2012} model for the shape and amplitude of the UV background. However, the relative abundance of carbon ions is sensitive to the details of the UV background model \citep{Finlator_2016, Keating_2016}. We have therefore also explored the effect of alternative UV background models on the \CII\ and \CIV\ fractions we predict for a given gas density and metallicity. We consider here the \citet{Puchwein_2019} and \citet{Faucher-Giguere_2020} models. The difference between these models and \citet{Haardt_2012} is an updated treatment of the IGM opacity during reionization, such that the models were designed to give a sensible evolution of the overall ionized hydrogen and helium fractions. They may therefore underestimate the number of photons with energies above the \ion{H}{i} and \ion{He}{i} ionization thresholds before hydrogen reionization is complete, and photons with energies above the \ion{He}{ii} ionization threshold before helium reionization is complete. For this reason we have used the \citet{Haardt_2012} model throughout this work, although as we will show all models produce similar results at fixed \ion{H}{i} column density at this redshift. However, we note that this similarity may be because we are considering models at $z=$~5, after hydrogen reionization has completed but before AGN dominate the UVB.

\begin{figure*}
	\includegraphics[width=2\columnwidth]{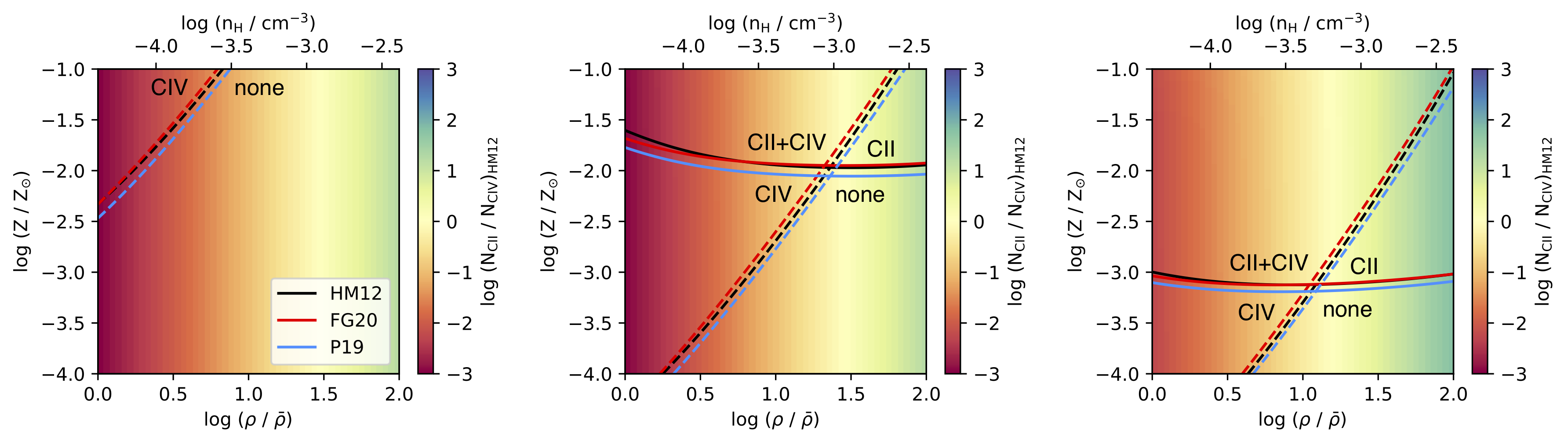}
    \caption{Effect of changing the UV background model on the detection regions for \ion{C}{ii} and \ion{C}{iv} absorbers. All models are shown at redshift 5. The different panels show 10\textsuperscript{15}~cm\textsuperscript{-2} (left),  10\textsuperscript{17}~cm\textsuperscript{-2} (middle) and 10\textsuperscript{19}~cm\textsuperscript{-2} (right). The black line is the \citet{Haardt_2012} model, the red line is the \citet{Faucher-Giguere_2020} model and the blue line is the \citet{Puchwein_2019} model. The background color map shows the N\textsubscript{CII}/N\textsubscript{CIV} ratio assuming the \citet{Haardt_2012} model for reference. As in Figure \ref{fig:modelling}, the marked quadrants indicate the detectability of \CII\ and \CIV\ absorption: \CII-only, no detection, \CIV-only, and \CII~+~\CIV.}
    \label{fig:Different_UVBs}
\end{figure*}

We repeat the \textsc{Cloudy} modelling as described in Section \ref{sec:modelling} and the results are shown in Figure \ref{fig:Different_UVBs}. We find that the positions of the \ion{C}{ii} and \ion{C}{iv} detection contours are relatively insensitive to the choice of UV background model. This is due to our choice to fix the \ion{H}{i} column density for a given hydrogen number density. This allows us to compensate with a change in the amplitude of the UVB with a change in the size of the absorber, to achieve a constant ionization parameter. Although these three models predict different amplitudes for the UV background at a given redshift, once the \ion{H}{i} column density is fixed, the differences in the intensity of the UV background as a function of wavelength become more relevant. Above redshift 4, all of these models are dominated by star-forming galaxies, so their shapes are relatively similar. However, the \textit{James Webb Space Telescope} has identified a population of faint AGN candidates into the EoR \citep[e.g.,][]{Harikane_2023}. If these faint AGN contribute to reionization \citep{Asthana2024}, this may lead to more hard \ion{C}{iv}-producing photons than are included in the UVBs investigated here, which could impact our modelled N\textsubscript{CII}/N\textsubscript{CIV} ratio.

\bsp	
\label{lastpage}
\end{document}